%% file: main.tex
\documentclass[sigconf]{acmart}

\AtBeginDocument{%
  }

\setcopyright{acmlicensed}
\copyrightyear{2018}
\acmYear{2018}
\acmDOI{XXXXXXX.XXXXXXX}
\acmConference[Conference acronym 'XX]{Make sure to enter the correct
  conference title from your rights confirmation email}{June 03--05,
  2018}{Woodstock, NY}
\acmISBN{978-1-4503-XXXX-X/2018/06}

\usepackage{tcolorbox}
\definecolor{lightgray}{HTML}{E8ECEF}
\usepackage{soul, color, xcolor}  
\usepackage{multirow}
\usepackage{array}
\usepackage[table]{xcolor}
\usepackage{pgfmath}
\usepackage{multirow}
\usepackage{pifont}
\usepackage{enumitem}

\usepackage{array,multirow,booktabs}   
\usepackage{graphicx} 
\usepackage{subcaption}

\begin{document}

\title{
Decoding Human-LLM Collaboration in Coding: An Empirical Study of Multi-Turn Conversations in the Wild
}


\author{Binquan Zhang, Li Zhang, Haoyuan Zhang, Fang Liu, Song Wang, Bo Shen, An Fu, Lin Shi}

\input{sec/0_abs}

\setcopyright{none} 
\settopmatter{printacmref=false} 


\maketitle


\input{sec/1_intro}

\input{sec/2_bg}

\input{sec/3_method}

\input{sec/4_pattern}

\input{sec/5_following}

\input{sec/6_human}

\input{sec/7_discussion}
\input{sec/8_conclusion}



\bibliographystyle{ACM-Reference-Format}
\bibliography{ref}

\end{document}

%% file: sec/0_abs.tex
\begin{abstract}
Large language models (LLMs) are increasingly acting as dynamic conversational interfaces, supporting multi-turn interactions that mimic human-like conversation and facilitate complex tasks like coding. While datasets such as LMSYS-Chat-1M and WildChat capture real-world user-LLM conversations, few studies systematically explore the mechanisms of human-LLM collaboration in coding scenarios. 
What tortuous paths do users experience during the interaction process? How well do the LLMs follow instructions?
Are users satisfied?
In this paper, we conduct an empirical analysis on human-LLM coding collaboration using LMSYS-Chat-1M and WildChat datasets to explore the human-LLM collaboration mechanism, LLMs' instruction following ability, and human satisfaction.
This study yields interesting findings:
1) Task types shape interaction patterns(linear, star and tree), with code quality optimization favoring linear patterns, design-driven tasks leaning toward tree structures, and queries preferring star patterns; 
2) Bug fixing and code refactoring pose greater challenges to LLMs' instruction following, with non-compliance rates notably higher than in information querying;
3) Code quality optimization and requirements-driven development tasks show lower user satisfaction, whereas structured knowledge queries and algorithm designs yield higher levels.
These insights offer recommendations for improving LLM interfaces and user satisfaction in coding collaborations, while highlighting avenues for future research on adaptive dialogue systems. We believe this work broadens understanding of human-LLM synergies and supports more effective AI-assisted development. We provide the data and scripts online to facilitate replication or future work at: \url{https://anonymous.4open.science/r/human-LLM-Eval}.

\end{abstract}

%% file: sec/1_intro.tex
\section{Introduction}




Large language models (LLMs) are transforming into versatile conversational agents capable of sustaining multi-turn dialogues, as seen in tools like ChatGPT~\cite{openai2024gpt4}, Copilot~\cite{github2023copilot}, Gemini~\cite{deepmind2023gemini}, and DeepSeek~\cite{deepseek2024coder}.
This evolution moves beyond isolated queries toward fluid, human-like exchanges that aid in intricate workflows ~\cite{jiang2024survey,jin2024llms,liu2024large}, including real-world issue resolution ~\cite{jimenez2023swe,mu2025experepair,xie2025swe}, automated testing ~\cite{alshahwan2024automated,deng2024pentestgpt} and debugging ~\cite{tian2024debugbench,epperson2025interactive,kang2025explainable,lyu2025automatic}, and full-cycle software development ~\cite{he2025llm}.
Yet, empirical evidence highlights persistent challenges in these interactions, particularly in coding domains where LLMs often falter in maintaining context over extended turns, leading to degraded performance and instruction deviations ~\cite{laban2025llms}.
For example, analyses of real-world datasets reveal that while multi-turn exchanges constitute about 73\% of user-LLM conversations ~\cite{wang2023mint}, models exhibit sharp drops in accuracy during prolonged dialogues ~\cite{laban2025llms}, with users frequently needing to correct errors or refine prompts to achieve viable outcomes ~\cite{zheng2024opencodeinterpreter}.
Such limitations underscore the need for deeper insights into interaction dynamics to mitigate context loss and enhance reliability in software engineering tasks.

Multi-turn interactions serve as a pivotal bridge in human-LLM collaboration, and analyzing interaction processes in real-world scenarios holds significant importance for enhancing user satisfaction and advancing LLMs ~\cite{southworth2023developing,horowitz2024adopting,brauner2023does}. Existing studies have curated and analyzed multi-turn interaction data from real-world settings. For instance, ~\citet{zheng2023lmsys} introduces the LMSYS-Chat-1M dataset, which encompasses one million conversations between real users and 25 LLMs. Similarly, ~\citet{zhao2024wildchat} examines user prompts directed at ChatGPT and releases the WildChat dataset, containing one million related conversations. 
While these datasets offer valuable resources for gaining insights into multi-turn interactions, few studies systematically dissect the mechanisms of human-LLM collaboration in real-world coding-related tasks, which is an essential step to unfold the complex, iterative problem-solving processes that characterize effective collaboration and to identify ways to enhance LLM-assisted coding workflows.
This raises intriguing questions: \textbf{\textit{What tortuous paths do users navigate when tackling coding tasks? How well do the LLMs follow instructions? Are users satisfied?}}

To bridge these gaps, we conducted an empirical analysis to decode the human-LLM coding collaboration and designed experiments to answer the following research questions:
\begin{itemize}[leftmargin=*]
    \item \textbf{RQ1 (Interaction Patterns):} What task types typically arise in human-LLM interactions, and how do these interactions proceed? This question examines task types within the interaction process, identifies underlying patterns, and provides theoretical foundations for optimizing interaction design.
    
    \item \textbf{RQ2 (LLM Instruction-Following Capabilities):} How effectively do LLMs follow instructions during human-LLM interactions? This question evaluates LLMs' capacity to comprehend and execute user directives, highlights their strengths and limitations, and informs strategies to enhance LLM performance and interaction efficiency.
    
    \item \textbf{RQ3 (Human Satisfaction):} What levels of satisfaction do humans experience in human-LLM interactions? This question assesses human satisfaction to support improvements in user experience and evaluation methodologies.

\end{itemize}

For our study, we first identify 5 primary task types in the human-LLM collaboration process (including design-driven development, requirements-driven development, code quality optimization, environment configuration, and information querying) along with 13 sub-task types (such as tool development, algorithm design and implementation, and bug fixing). We also uncover three main interaction patterns (linear, star, and tree). Second, we apply a variety of analysis techniques: (1) open card sorting \cite{rugg1997sorting} to examine the task types involved in the human-LLM interaction process; (2) Kruskal-Wallis H test \cite{kruskal1952use} and Kolmogorov-Smirnov (KS) test \cite{an1933sulla} to verify the significant impacts of task types, instruction-following capabilities, and interaction patterns; (3) satisfaction trajectory method \cite{smirnov1948table} to characterize changes in user satisfaction during the interaction process; (4) Cohen's Kappa \cite{cohen1960coefficient} to assess the consistency of classification labeling behaviors. 

Among these results, we find that:

\begin{itemize}
    \item Code quality optimization favors the linear pattern to enable incremental refinements, whereas design-driven development leans toward the tree pattern for deeper explorations, facilitating effective handling of complex code generation.
    
    \item  Bug fixing and code refactoring pose greater challenges to LLMs' instruction following, with non-compliance rates notably higher than in information querying.
    
    \item Code quality optimization and requirements-driven development tasks show lower user satisfaction, whereas structured knowledge queries and algorithm designs yield higher levels.
    Moreover, satisfaction declines as conversations lengthen, with task focus increasingly shifting toward error correction.
    
\end{itemize}

%% file: sec/2_bg.tex
\section{Related Work}


\subsection{Multi-Turn Interaction Evaluation}

With the rapid growth of ChatGPT sparking heightened focus on multi-turn assessments, early notable initiatives like MT-Eval ~\cite{kwan2024mt}  incorporated authentic dialogue patterns into GPT-4 for generating extended conversations.
Subsequent research built upon such benchmarks, for example by introducing finer-grained hierarchies for capability evaluation~\cite{bai2024mt}, modeling conversational logic through defined dialogue relations~\cite{li2025structflowbench}, addressing fairness in ongoing exchanges~\cite{fan2024fairmt}, or training specialized models to emulate user-driven multi-turn queries~\cite{sun2024parrotenhancingmultiturninstruction}. Other efforts simulated specific scenarios, such as programmer-AI partnerships~\cite{mozannar2023simulating}, probed LLMs on intertwined instructions across difficulty and memory demands~\cite{han2025can}, or fragmented prompts to test recovery in incomplete dialogues~\cite{laban2025llms}, highlighting persistent challenges in maintaining coherence once models deviate.
Beyond relying on LLMs to mimic real-user dialogue behaviors for assessing LLM performance, several efforts have turned to gathering interaction data from genuine conversational contexts.
For instance, ~\citet{zheng2023lmsys} collected the LMSYS-Chat-1M dataset, which compiles one million authentic conversations from over 210K unique IP addresses, engaging 25 LLMs across more than 150 languages and a wide range of topics. 
Similarly, ~\citet{zhao2024wildchat} compiled over one million user-ChatGPT dialogues, encompassing more than 2.5 million interaction rounds.

While existing efforts predominantly rely on simulated environments or controlled datasets to assess LLM performance in multi-turn interactions, we emphasize empirical analysis of real-world human-LLM collaborations in coding tasks, drawing from large-scale datasets like LMSYS-Chat-1M and WildChat. Rather than focusing solely on model capabilities or methodological enhancements, we dissect interaction patterns, task types, and user satisfaction dynamics to uncover practical insights into collaboration mechanisms, including the impacts of instruction compliance and evolving user experiences.

\subsection{LLM-based Coding}

LLMs have achieved substantial progress in supporting code-related activities, with a growing emphasis on enabling effective human-LLM interactions during coding workflows. 
Numerous models tailored for collaborative coding have emerged in recent years~\cite{chen2021codex, fried2022incoder, li2023starcoder, li2022alphacode, nijkamp2022codegen, nijkamp2023codegen2, roziere2023codelama,geminiteam2023gemini, anthropic2024claude3, guo2024deepseekcoder,openai2024gpt4}. Codex~\cite{chen2021codex}, an early milestone, leveraged generative pre-training on up to 12 billion parameters to produce code snippets, powering Copilot's real-time assistance and transforming how developers engage with AI in iterative coding sessions. This breakthrough spurred further developments: DeepMind's AlphaCode~\cite{li2022alphacode} focused on competitive programming scenarios that mimic human problem-solving dialogues; Meta released InCoder~\cite{fried2022incoder} and Code Llama~\cite{roziere2023codelama} to facilitate more contextual exchanges; the BigCode project delivered StarCoder~\cite{li2023starcoder} for open-source collaboration; and OpenAI's GPT series, including GPT-4~\cite{openai2024gpt4}, refined via Reinforcement Learning from Human Feedback (RLHF), excelled in sustaining natural, turn-based coding conversations. More recent advances include Google's Gemini~\cite{geminiteam2023gemini}, which supports multimodal inputs for richer developer-LLM interactions; Anthropic's Claude models~\cite{anthropic2024claude3}, emphasizing safety and extended context for reliable multi-step coding guidance; and DeepSeek's DeepSeek-Coder~\cite{guo2024deepseekcoder}, optimized for programming languages to enhance precision in iterative refinement during collaborative tasks.

These advancements have not only boosted generation accuracy but also enabled more fluid human-LLM partnerships, where users guide models through clarifications, revisions, and explorations in real-time coding sessions.

%% file: sec/3_method.tex
\section{Methodology and Study Design}

\begin{figure*}[t!]
\centering
\setlength{\abovecaptionskip}{0.1cm}
\includegraphics[width=0.95\textwidth]{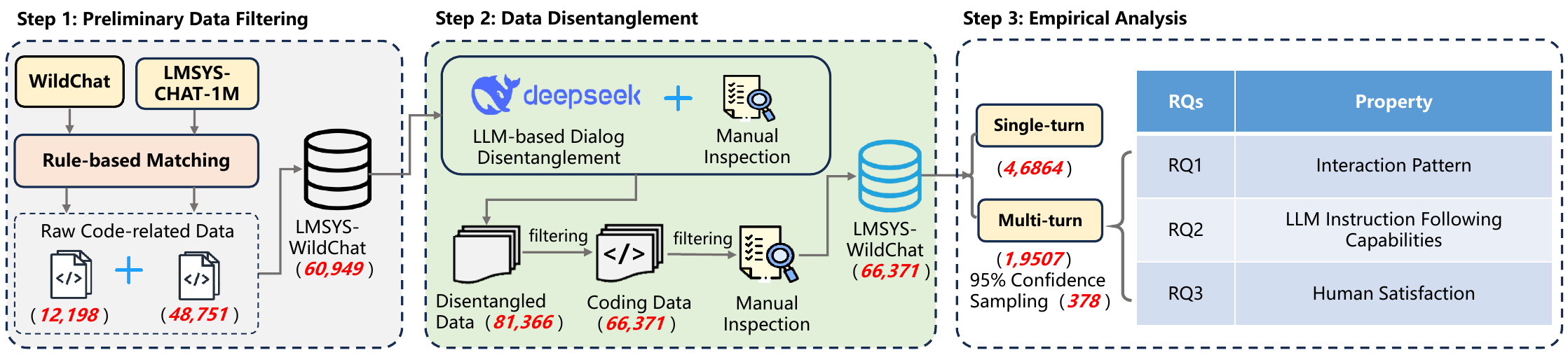}
\caption{Overview of this study}
\label{fig_method}
\vspace{-0.35cm}
\end{figure*}

Overall, the research methodology comprises three phases, as illustrated in Figure ~\ref{fig_method}. First, in the initial data screening phase, we selected two publicly available datasets that capture real-world human-LLM interaction scenarios. We then applied a rule-based matching approach to extract code-related conversation logs.
Second, in the data decoupling phase, we employed an LLM-based method to automatically disentangle the interaction data, followed by manual verification to ensure decoupling accuracy. 
From the disentangled data, we further filtered coding-related dialogues and conducted a manual review, yielding 66,371 entries encompassing both single-turn and multi-turn conversations. 
Finally, in the empirical analysis phase, we sampled 378 multi-turn dialogues from the decoupled dataset (19,507 in total) at a 95\% confidence level (with a ±5\% margin of error) to examine human-LLM multi-turn interaction patterns, instruction following, and human satisfaction, thereby addressing our research questions more effectively.

\subsection{Preliminary Data Filtering}
The data for this study primarily originates from two publicly available datasets that capture real-world human-LLM interactions: LMSYS-Chat-1M ~\cite{zheng2023lmsys} and WildChat~\cite{zhao2024wildchat}. 
They span a broad array of topics, such as software design, programming, cultural, social, and geographical discussions, reflecting the diverse nature of human-LLM interactions in real scenarios. Observations reveal that both datasets involve multiple programming languages, including Python, Java, Rust, and Go. 
To extract coding-related tasks from these datasets, we designed and applied a rule-based matching approach. 
Specifically, we referenced the top 20 programming languages listed in the TIOBE index~\cite{tiobe}, integrating their full spellings (e.g., Python) along with common abbreviations and file extensions (e.g., .py) to craft targeted and exhaustive matching criteria. 
This process yielded 48,751 coding-related interactions from LMSYS-Chat-1M and 12,198 from WildChat.
Given the consistent format and organization of both datasets, and considering that WildChat primarily focuses on interactions with GPT-3.5-Turbo and GPT-4, complemented with LMSYS-Chat-1M, we merged the extracted data into a unified dataset named LMSYS-WildChat (60,949) to facilitate subsequent processing and analysis.

\subsection{Data Disentanglement}

Preliminary analysis of the LMSYS-WildChat dataset (60,949) reveals that individual conversations often encompass multi-topic tasks, along with numerous interactions unrelated to coding. 
To derive single-topic interactions focused exclusively on coding—ensuring each resulting conversation addressed only one cohesive theme, though potentially encompassing subtasks within that theme—we leveraged few-shot prompting with an LLM (DeepSeek-V3-0324) to disentangle the data. 
Following guidelines from the DeepSeek documentation~\cite{deepseek-params}, we configured the temperature at 1.0 while keeping other parameters at their defaults to promote reliable and consistent outputs, resulting in 81,366 disentangled entries.
To validate the effectiveness of this process, we randomly sampled 383 entries (at a 95\% confidence level with a ±5\% margin of error) and had two annotators independently evaluate them against the original conversations, focusing on whether the LLM accurately isolated single-topic segments. 
The annotators reached a Cohen's Kappa of 0.87, reflecting strong agreement and, by extension, confirming the LLM's effectiveness in this task, with an overall accuracy of 92\% based on consensus labels.
We then applied the same rule-based filtering as in the initial preprocessing to retain only those decoupled conversations pertaining to programming tasks, yielding 66,371 conversation logs: 46,864 single-turn and 19,507 multi-turn. 
For an in-depth examination of multi-turn human-LLM interaction patterns, instruction following, and human satisfaction, we selected 378 samples from the multi-turn set (again at a 95\% confidence level with a ±5\% margin of error). 
To confirm their relevance, we manually reviewed these 378 samples, verifying that all were indeed coding-focused, with no discrepancies identified.

%% file: sec/4_pattern.tex
\section{RQ1: Interaction Pattern}
\subsection{Methodology}
We identified the task types and instruction intents in multi-turn conversation data using an open card sorting method \cite{rugg1997sorting}. 
Specifically, we formed an expert analysis team and a student annotation team. 
Our expert analysis team includes four PhDs in Software Engineering or Computer Science, each with over 10 years of development experience (Java, Python) and extensive research expertise in intelligent software engineering and LLMs. The student annotation team comprises four graduate students in intelligent software engineering, whose 3+ years of development experience ensures that they can accurately interpret user and developer instructions for the annotation task.

\textbf{Step 1: Exploratory analysis by the expert team.} 
The experts randomly selected 80 conversations from the multi-turn conversation samples (378) and conducted exploratory analysis using open card sorting. Each conversation serves as an independent "card." 
The four experts were divided into two random groups and thoroughly examined each conversation to determine its primary task type, the user's instruction intent, and the interaction pattern. 
Since the data has been decoupled, each conversation was reduced to a single task type. For any remaining conversations with multiple task types, we retained only the dominant one to ensure consistency and comparability.
Our initial observations indicate that user instructions within conversations typically exhibit a single intent. 
Additionally, to identify interaction patterns, the experts mapped the topological structure of each dialogue by representing it as a graph: excluding the root (Virtual Node), each node denoted a conversational turn (encompassing both user prompt and LLM response), with directed edges indicating dependencies between turns. This mapping allowed classification into one of three patterns—linear (sequential progression), star (central theme with independent branches), or tree (hierarchical expansion with sub-branches)—based on the resulting structure.
During analysis, experts engaged in thorough discussions to resolve any disagreements until a consensus was reached. 
Ultimately, they identified 5 task types (including 12 subtypes), as show in Table ~\ref{rq_1_task_type}, 11 instruction intents and 3 primary interaction patterns, as shown in Figure ~\ref{rq1_pattern}.
Among them, the instruction intents are: add new feature(ANF), fix bugs(FB), refactor code(RC), clarify requirements(CR), review code(RevC), Information query(IQ), request code implementation(RCI), question LLM's response(QLR), confirmation /continuation request(CCR), polite greeting(PG), others.
Due to space constraints, it is available in our supplementary data repository (see data link).
The analysis process showed high consistency, with average Cohen's Kappa values of 0.78 or greater.


\textbf{Step 2: Annotation by the student team.} 
Building on the task types derived from the exploratory analysis, the student annotation team independently annotated the remaining 298 samples. 
The four graduate students were equally divided into two groups. Annotators in each group individually assessed each card's content and assigned it a task type.
We then collected the results from both groups for initial statistical analysis. Disagreements in classification were resolved through moderated group discussions until a consensus was reached. 
If a potential new task type emerged, the student team consulted with the expert analysis team to confirm its validity and inclusion.
While the process accommodated such possibilities, no additional categories were identified beyond those from the initial expert phase.
To ensure annotation quality, we evaluated inter-annotator reliability by calculating Cohen's Kappa for each group. The average Kappa was 0.85, indicating substantial agreement. This systematic process yielded a final set of 5 task types (with 12 subtypes), as show in Table ~\ref{rq_1_task_type}, 11 instruction intents(see data link), and 3 primary interaction patterns, as detailed in Figure ~\ref{rq1_pattern}.

\input{tab/rq_1_task_type}

\input{}

\subsection{Result and analysis}


\subsubsection{Interaction Patterns.}
\begin{figure}
    \centering
    \setlength{\abovecaptionskip}{0.1cm}
    \includegraphics[width=0.45\textwidth]{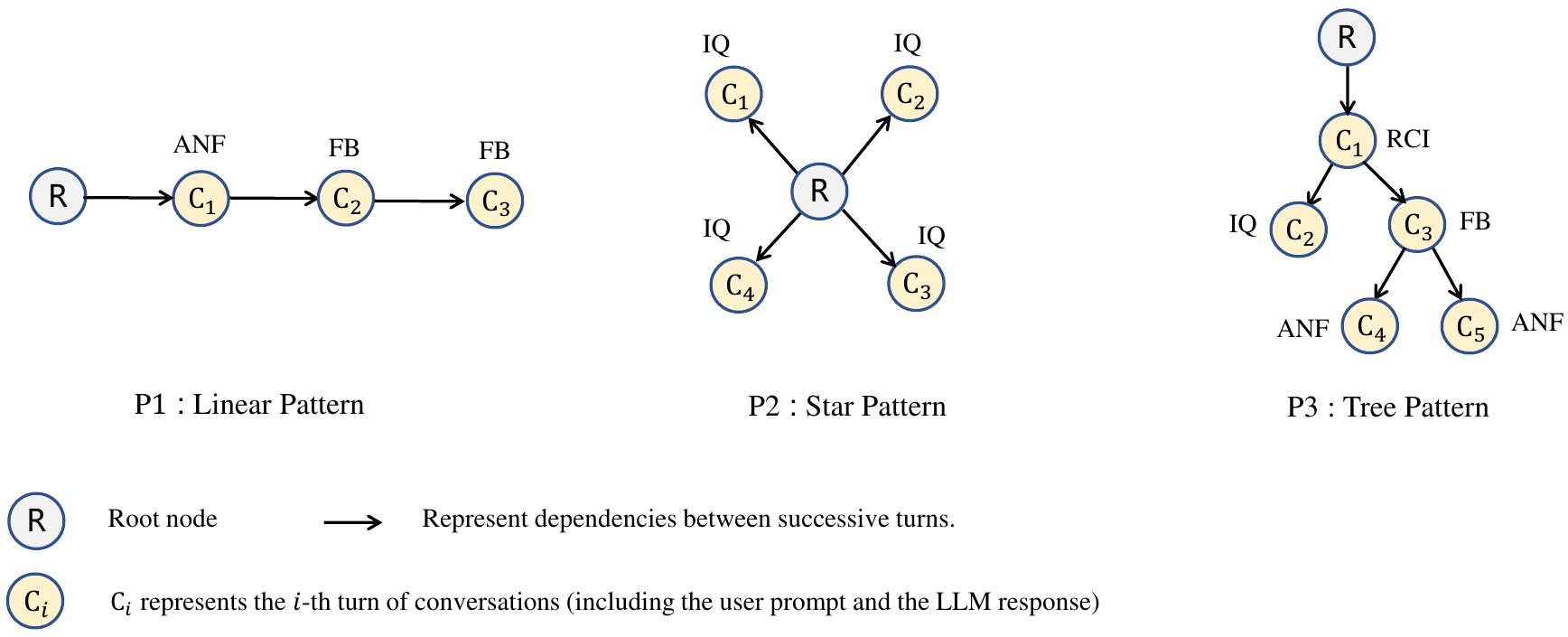}
    \caption{Interaction Patterns} 
    \label{rq1_pattern}
    \vspace{-0.68cm}
\end{figure}

We have identified three primary interaction patterns, as illustrated in Figure  ~\ref{rq1_pattern}. 
Excluding the root, each node represents a conversational turn (including both user prompt and LLM response). Arrows denote dependencies between successive turns.
\textbf{(1) P1: Linear Pattern (65.87\%)} advances sequentially, with each round building directly on the output of the previous round to ensure coherent information and logical flow.
\textbf{(2) P2: Star Pattern (15.61\%)} centers on a core theme or root node, from which multiple independent branches derive, each maintaining a direct association with the center.
\textbf{(3) P3: Tree Pattern (18.52\%)} originates from an initial theme or root node and progressively expands into multiple branches and sub-branches, forming a hierarchical structure; each branch path constitutes an independent exploration direction, with the overall structure relying on outputs from upper-level nodes.


\subsubsection{Tasks and Patterns}
To uncover users' preferences for interaction patterns across different task requirements, we explored the associations between task types and interaction patterns based on the results of Fisher's exact test and standardized residuals analysis.
First, Fisher's exact test produces a p-value of 0.0016 (<0.05), offering sufficient statistical evidence to reject the null hypothesis.
We then analyzed the standardized residuals to determine specific relationships (see Table ~\ref{rq_1_std}), where absolute values greater than 1.96 indicate significance and greater than 1.0 indicates a tendency.
Specifically, code quality optimization significantly avoids the star pattern (residual = -2.68) while favoring the linear pattern (residual = 1.39), as its systematic, incremental nature suits contextual tasks like code refactoring over the star pattern's divergent explorations. 
Similarly, query information tasks showed a strong preference for the star pattern (residual = 2.520), as its divergent structure is ideal for exploratory exchanges, such as branching from a core topic to retrieve information from multiple perspectives.
Other tasks showed notable, though not significant, tendencies (|residual| > 1.0). 
Design-driven development favored the tree pattern (residual = 1.561) while avoiding the star pattern (residual = -1.222), as the tree’s branching structure is well-suited for iterative problem breakdown in complex design. 
In contrast, environment configuration tasks leaned toward the star pattern (residual = 1.504), likely because its divergent trials are effective for troubleshooting a central issue. 
Finally, requirement-driven development showed no clear pattern preference (all |residuals| < 1.0), suggesting greater interaction flexibility.

\input{tab/rq_1_std}

\begin{tcolorbox}[colback=lightgray, colframe=lightgray, boxrule=0.5pt, arc=5pt,left=5pt,right=5pt,top=5pt,bottom=5pt]
\textbf{Finding 1:}
Task types significantly influence the choice of interaction patterns. 
Specifically, code quality optimization tasks tend to avoid the star pattern and depend more heavily on the linear pattern. 
In contrast, system-level development tasks favor the tree pattern. Meanwhile, knowledge query tasks and environment configuration tasks lean towards the star pattern.

\end{tcolorbox}

\subsubsection{Turns and Task Impact on Patterns}
To capture how users shape interaction patterns across varying conversation lengths and task demands, we further investigate the influence of interaction turns and task types on interaction patterns. 

First, we observed that low-turn interactions ($\leq$ 4 turn) tend to favor simpler linear patterns (72.40\%), while high-turn interactions (> 4 turns) tend to favor complex tree patterns (61.43\%). Therefore, we consider 4 rounds as a key transition point where the interaction pattern gradually shifts from a simple linear structure to a more branching tree structure.
Based on this, we conducted a stratified analysis of interaction pattern distributions by task type and turn length (Low: $\leq$4 turns; High: >4 turns), as presented in Table ~\ref{rq_1_fenceng}.
Specifically, under low turns, all task types predominantly feature the linear pattern, with proportions ranging from 52.9\% (environment configuration) to 86.8\% (code quality optimization). 
For code quality optimization tasks, the linear pattern accounts for 86.8\%, star for 5.9\%, and tree for 7.4\%. 
This arises because the initial stages of code optimization often focus on localized fixes, allowing efficient iterative validation without introducing branching paths. 
Similarly, requirement-driven development exhibits a linear proportion of 74.7\%. This likely stems from straightforward user requirements, such as generating short code snippets, which a linear progression adequately addresses. 
In contrast, the star pattern appears more prominent in information query (28.7\%) and environment configuration (35.3\%) tasks, indicating a tendency for divergent exploration in short interactions, such as multi-dimensional knowledge queries. 
The tree pattern generally accounts for 13\% or less across task types, a finding that underscores the preference for simpler structures in constrained conversations. 
This low prevalence further illustrates that in short interactions, users gravitate toward direct, linear progression as a means to swiftly accomplish their objectives.

The distribution shifts markedly under high turns, with the tree pattern dominating all tasks at proportions from 50.0\% (code quality optimization) to 75.0\% (environment configuration). In design-driven development, the tree pattern accounts for 64.7\%, while the linear pattern drops to 29.4\%.
This reflects the nature of design-driven development, which often starts with high-level design and proceeds to modular implementation via branching paths.
In code quality optimization, the tree pattern reaches 50.0\%, primarily because it supports the construction of deep debugging chains. By comparison, information query tasks see the tree pattern surge to 61.5\%. 
Notably, as turns increase, patterns evolve from star-like divergent structures to integrated tree structures. This suggests that persistent information query dialogues often require nested sub-questions to build a comprehensive understanding.
Correspondingly, the star pattern nearly disappears (mostly 0\%), while the linear pattern, though declining to 25.0\%–50.0\%, still maintains a notable presence. This indicates that even in deep information-seeking conversations, some shallow explorations continue in a linear manner, particularly in information query tasks.

\input{tab/rq_1_fenceng}

\begin{tcolorbox}[colback=lightgray, colframe=lightgray, boxrule=0.5pt, arc=5pt,left=5pt,right=5pt,top=5pt,bottom=5pt]
\textbf{Finding 2:}
Interaction turns and task types collectively drive the evolution of interaction patterns.
In low-turn interactions (turns $\leq$ 4 ), all tasks predominantly adopt the linear pattern, with code quality optimization showing the highest proportion (86.8\%). The star pattern stands out in environment configuration (35.3\%) and information query (28.7\%) tasks.
In high-turn interactions (turns > 4), all tasks shift to the tree pattern as the dominant form, with environment configuration reaching the peak (75.0\%); the linear pattern decreases to 25.0\%-50.0\%.

\end{tcolorbox}
\vspace{-0.4cm}

%% file: tab/rq_1_task_type.tex
\begin{table}[t!]
\setlength{\abovecaptionskip}{0.1cm}
\caption{Task Types and Subtasks in Human-LLM Coding Collaboration}
\scriptsize
\renewcommand{\arraystretch}{1.1} 
\setlength{\tabcolsep}{1.5pt} 
\renewcommand{\multirowsetup}{\centering}

\begin{tabular}{
    >{\centering\arraybackslash}m{1.7cm} 
    >{\centering\arraybackslash}m{1.9cm} 
    >{\raggedright\arraybackslash}m{3.6cm} 
    >{\centering\arraybackslash}m{0.8cm} }
\hline
\textbf{Task Type} & \textbf{Subtask Type} & \textbf{Description} & \textbf{Num} \\ 
\hline

\multirow{3}{1.8cm}[-8ex]{Design-Driven Development} 
    & Tool Development 
    & Users request LLMs to design software or tools, encompassing architecture design, functional decomposition, and implementation details, etc. & 39 \\ 
\cline{2-4}
& Algorithm Design and Implementation 
    & Users request LLMs to design and implement algorithms, including steps, pseudocode, or full implementations for problem-solving, etc. & 10 \\ 
\cline{2-4}
& Data Science and Machine Learning 
    & Users request LLMs to design, modify, or optimize machine learning models, including feature engineering, architecture adjustments, etc. & 14 \\ 
\hline

\multirow{2}{1.8cm}[-2ex]{Requirement-Driven Development} 
    & Code Generation 
    & Users request LLMs to generate code from scratch, producing snippets or programs based on specified requirements. & 56 \\ 
\cline{2-4}
& Functional Expansion 
    & Users request LLMs to extend existing code by adding new features. & 41 \\ 
\hline

\multirow{2}{1.8cm}[-2ex]{Code Quality Optimization} 
    & Bug Fixing 
    & Users request LLMs to identify and fix bugs. & 67 \\ 
\cline{2-4}
& Code Refactor 
    & Users request LLMs to refactor code to eliminate smells, improve readability, or optimize performance, etc. & 23 \\ 
\hline

\multirow{2}{1.8cm}[-2.5ex]{Environment Configuration} 
    & Environment Configuration and Deployment 
    & Users request LLMs to configure environments and guide deployment, including dependencies and platforms, etc. & 11 \\ 
\cline{2-4}
& Configuration Debug 
    & Users request LLMs to troubleshoot configuration issues, such as settings, environmental mismatches, etc. & 10 \\ 
\hline

\multirow{3}{1.8cm}[-5.5ex]{Information Query} 
    & Programming Knowledge Query 
    & Users query LLMs for programming knowledge, including explanations of built-in methods, syntax, etc. & 35 \\ 
\cline{2-4}
& Code Explanation 
    & Users request LLMs to explain code segments, covering functional modules, logic flows, or specific operations, etc. & 59 \\ 
\cline{2-4}
& Database Knowledge Query 
    & Users query LLMs for database-related knowledge, such as operations in systems like MySQL, query optimization, or schema design principles, etc. & 13 \\ 
\hline
\end{tabular}
\label{rq_1_task_type}
\vspace{-0.5cm}
\end{table}

%% file: tab/rq_1_std.tex
\begin{table}[t]
\caption{Standardized Residuals for Task Types and Interaction Patterns} 

\centering
\begin{tabular}{lccc}
\toprule
\multicolumn{1}{c}{Task Type} & \multicolumn{1}{c}{Linear} & \multicolumn{1}{c}{Star} & \multicolumn{1}{c}{Tree} \\
\midrule
Design-Driven Development & -0.233 & -1.222 & 1.561 \\
Requirement-Driven Development & 0.263 & 0.221 & -0.699 \\
Code Quality Optimization & 1.392 & -2.681 & -0.163 \\
Environment Configuration & -1.031 & 1.504 & 0.563 \\
Information Query & -0.891 & 2.520 & -0.632 \\
\bottomrule
\end{tabular}

\label{rq_1_std} 
\label{rq_3_manyidu}
    \vspace{-0.4cm}
\end{table}

%% file: tab/rq_1_fenceng.tex
\begin{table}[t!]
\setlength{\abovecaptionskip}{0.1cm}
\caption{Distribution of Interaction Patterns by Task Type and Turn Length}
\footnotesize 
\renewcommand{\arraystretch}{1.1}
\setlength{\tabcolsep}{1.5pt}

\begin{tabular}{ 
    >{\raggedright\arraybackslash}m{2.5cm} 
    >{\centering\arraybackslash}m{0.8cm} 
    >{\centering\arraybackslash}m{0.8cm} 
    >{\centering\arraybackslash}m{0.8cm} 
    >{\centering\arraybackslash}m{0.8cm} 
    >{\centering\arraybackslash}m{0.8cm} 
    >{\centering\arraybackslash}m{0.8cm} }
\toprule
\textbf{Task Type} & \multicolumn{3}{c}{\textbf{Low ($\leq$4)}} & \multicolumn{3}{c}{\textbf{High ($>$4)}} \\ 
\cmidrule(lr){2-4} \cmidrule(lr){5-7}
 & Linear(\%) & Star(\%) & Tree(\%) & Linear(\%) & Star(\%) & Tree(\%) \\ 
\midrule
Code Quality Optimization & 86.8 & 5.9 & 7.4 & 50.0 & - & 50.0 \\ 
Design-Driven Development & 76.1 & 10.9 & 13.0 & 29.4 & 5.9 & 64.7 \\ 
Environment Configuration & 52.9 & 35.3 & 11.8 & 25 & - & 75 \\ 
Information Query & 61.7 & 28.7 & 9.6 & 38.5 & - & 61.5 \\ 
Requirement-Driven Development & 74.7 & 19.3 & 6.0 & 28.6 & - & 71.4 \\ 
\bottomrule
\end{tabular}
\label{rq_1_fenceng}
\vspace{-0.4cm}
\end{table}

%% file: sec/5_following.tex
\section{RQ2: LLM Instruction Following Capabilities}

\subsection{Methodology}

To evaluate the instruction-following ability of LLMs in human-LLM interactions, we adopted a dual evaluation mechanism that assesses LLM responses in multi-turn conversations.
Given the large scale and diversity of interactions in the WildChat and LMSYS-Chat-1M datasets, we employ LLM-based automatic evaluation rather than predefined test cases to enhance scalability.
Specifically, we first selected DeepSeek-Reasoner \cite{guo2025deepseek} as the primary evaluator, which has acquired advanced reasoning abilities through reinforcement learning, including features such as self-verification and reflection. It excels in complex instruction evaluation tasks and handles multi-layered instruction analysis effectively. 
We provide DeepSeek-Reasoner with custom prompts (details in the data link), incorporating full dialogue context, user instructions, and LLM responses, to evaluate turn-by-turn whether each response fully complies with the instructions.
Default parameter settings were employed to optimize evaluator performance.

Second, to ensure result reliability, we conducted manual sampling validation on DeepSeek outputs. The sampling covers 20\% of the data and spans all five task types. Validators consist of two experts from the expert team and two graduate students from the annotation team. All possess extensive experience in LLM interaction analysis. The validation process follows an independent assessment protocol. Each validator reviewed samples individually and assigned instruction-following classifications. For samples with discrepancies, we resolved them via majority voting to minimize subjective bias. Ultimately, manual validation results align with DeepSeek outputs at over 85\% consistency. This confirms the automated evaluator's effectiveness and bolsters the credibility of our experimental conclusions.

Finally, we adopted two metrics from MultiIF~\cite{he2024multi}: instruction-level loose accuracy and conversation-level loose accuracy to quantify the instruction-level and conversation level following capabilities of the LLMs.
Instruction-level loose accuracy measures the consistency of individual LLM responses to user instructions under a relaxed criterion. 
It handles introductory phrases, such as ``Certainly, here is the revised version''.
Conversation-level loose accuracy assesses overall dialogue compliance between users and LLMs using the same relaxed standard.

\subsection{Result and analysis}



Our evaluation reveals a conversation-level loose accuracy of 24.07\% and an instruction-level loose accuracy of 48.24\%. In this section, we provide a detailed analysis from both conversation-level and instruction-level perspectives.

\subsubsection{Conversation-level following capability}

The conversation-level loose accuracy is only 24.07\%. 
This reveals that roughly 75.93\% of conversations in the dataset include at least one instance of non-compliance. To investigate the drivers of LLMs' limited instruction-following ability, we examined the data through the lenses of task types and interaction patterns.


\textbf{Tasks and Instruction Non-Compliance.}
To examine whether task types significantly influence LLMs' instruction non-compliance, we apply the Kruskal-Wallis H test given the non-normal distribution of the non-compliance data.
Results yield a p-value of 0.6658 (> 0.05). 
Although no statistically significant differences emerged across high-level task categories, systematic variations could still arise at the subtask level. 
We therefore conducted a finer-grained analysis to evaluate how specific subtask types affect non-compliance, once again applying the Kruskal-Wallis H test to non-parametrically compare median differences in instruction non-compliance across these subtasks.
Experimental results indicate that the p-value of 0.04 (< 0.05) reflects a statistically significant impact of subtask types on failure to follow instructions in LLMs.
The significance implies that variation in LLM performance is driven primarily by subtask-specific factors rather than general task categories.

\begin{figure}
    \centering
    \setlength{\abovecaptionskip}{0.1cm}
    \includegraphics[width=0.45\textwidth]{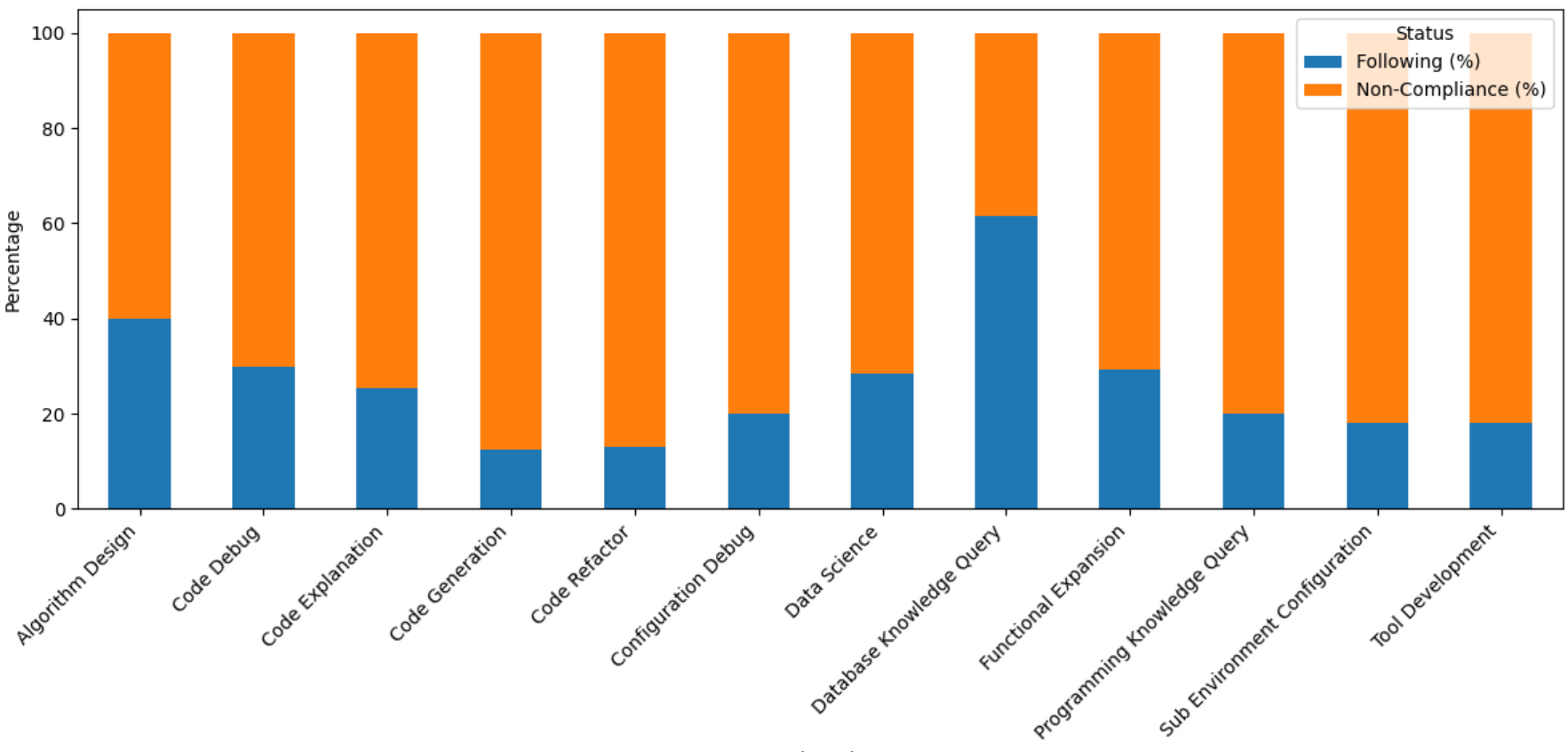}
    \caption{Distribution of Instruction Following and Non-Compliance Across Different Subtask Types} 
    \label{subtask_non}
    \vspace{-0.4cm}
\end{figure}

A closer examination of adherence rate distributions across subtasks, as illustrated in Figure ~\ref{subtask_non}, reveals substantial variability in non-compliance rates. 
Notably, the database knowledge query subtask shows the lowest non-compliance rate (38.46\%), which indicates LLM's strong instruction compliance in handling structured queries. This likely stems from the deterministic nature of such tasks and the standardized aspects of knowledge retrieval. In contrast, code generation, code refactoring, tool development, and environment configuration and deployment exhibit the highest non-compliance rates, at 87.50\%, 86.96\%, 82.05\%, and 81.82\%, respectively. These tasks typically involve intricate generation or modification processes, which heighten the risk of instruction non-compliance.

Other subtasks, including code explanation (74.58\%), data science and machine learning (71.43\%), bug fixing (70.15\%), and functional expansion (70.73\%), display intermediate non-compliance rates, reflecting moderate challenges. For instance, bug fixing often requires iterative repairs, while data science and machine learning tasks entail modifications to complex models. These differences underscore that, in coding-related interactions, generation and modification subtasks more readily provoke instruction non-compliance than query-based subtasks. Such insights guide future efforts to enhance LLMs' multi-turn instruction-following capabilities.

\begin{tcolorbox}[colback=lightgray, colframe=lightgray, boxrule=0.5pt, arc=5pt,left=5pt,right=5pt,top=5pt,bottom=5pt]
\textbf{Finding 3:}
Code generation, code refactoring, tool development, and environment configuration and deployment commonly provoke non-compliance, with rates exceeding 80\%. Code explanation, data science and machine learning, bug fixing, and functional expansion show comparable challenges, exhibiting rates of failure to follow instructions of 70\% to 75\%.

\end{tcolorbox}

\textbf{Interaction Patterns and Instruction Non-Compliance.}
To further examine the influence of interaction patterns on instruction non-compliance, we applied the Kruskal-Wallis H test to compare non-compliance rates across patterns (p = 1.265e-04).
These results indicate that interaction structure non-randomly influences LLM's instruction adherence.
Building on this, we computed statistics on instruction non-compliance under each pattern. As shown in Table ~\ref{rq_2_std}, the tree pattern exhibits the highest rate (94.28\%), showing that LLMs in tree interactions almost invariably omit or deviate from at least one instruction. The linear pattern follows at 73.49\%, while the star pattern shows the lowest rate (64.40\%). This distribution implies that the star pattern, despite its multi-branching nature, benefits from a structure centered on a core topic, which may ease LLMs' context management. In contrast, the deep branching in the tree pattern markedly amplifies challenges.


\input{tab/rq_2_std}

\begin{tcolorbox}[colback=lightgray, colframe=lightgray, boxrule=0.5pt, arc=5pt,left=5pt,right=5pt,top=5pt,bottom=5pt]
\textbf{Finding 4:}
The tree pattern emerges as a "high-risk" pattern for failures to follow instructions. 
Its non-compliance rate significantly exceeds that of the other two (p < 0.05). In contrast, the linear and star patterns show no significant differences.

\end{tcolorbox}

\subsubsection{Instruction-level following capability}
At the instruction level, loose accuracy reaches 48.24\%, implying that over half (51.76\%) of individual instructions encounter some form of non-compliance. 
In exploring the underlying factors constraining LLMs' following of instructions, we analyzed the data with a focus on turns and instruction intents.

\textbf{Turns and Instruction Non-Compliance}
To uncover the spatiotemporal distribution patterns of instruction non-compliance in human-LLM interactions, we investigated the positional preferences for such occurrences, specifically whether they tend to emerge in the early or late stages of conversations.

For short conversations (turn $\leq $4), given their limited and fixed number of turns, we employed absolute position statistics (positions 1 through 4) to precisely capture early deviations. For long conversations (turn > 4), which exhibit substantial variability in turn counts (ranging from 5 to dozens), we adopted a relative position binning strategy. This approach normalizes dialogue length into five equal-width intervals: Early (1-20\%), Early-Mid (21-40\%), Mid (41-60\%), Mid-Late (61-80\%), and Late (81-100\%), thereby mitigating the influence of absolute turn differences. 
We then applied the Kolmogorov–Smirnov (KS) test to assess whether the distribution of non-compliance positions deviates from uniformity. A p-value below 0.05 indicates a significant deviation, suggesting positional bias.

As shown in Table ~\ref{rq_2_turn}, short conversations exhibit a highly uneven distribution of non-compliance, concentrated primarily in the early stages: position 1 accounts for 38.68\%, position 2 peaks at 41.22\%, position 3 drops to 14.25\%, and position 4 stands at only 5.85\%. The KS test yields 1.4088e-31 (far below 0.05), confirming significant non-uniformity. In long conversations, the distribution is comparatively flatter yet clearly skewed toward later stages: the Early interval comprises 16.48\%, Early-Mid 18.35\%, Mid 17.98\%, Mid-Late 19.85\%, and Late reaching the highest at 27.34\%. The KS test produces a p-value of 5.7985e-06 (p < 0.05), likewise endorsing non-uniform distribution, with the latter 40\% cumulatively representing 47.19\%, underscoring late-stage clustering.

\input{tab/rq_2_turn}

\begin{tcolorbox}[colback=lightgray, colframe=lightgray, boxrule=0.5pt, arc=5pt,left=5pt,right=5pt,top=5pt,bottom=5pt]
\textbf{Finding 5:}
In short conversations, instruction non-compliance commonly occurs at turn 2. In long dialogues, instruction non-compliance tends to occur in the late portion (i.e., 81-100\% of total turns), which corresponds to the last 20\%.

\end{tcolorbox}

\textbf{Instruction Intent and Instruction Non-Compliance }

To investigate the impact of instruction intent types on LLMs' ability to follow instructions, we hypothesize that different intent types lead to significant variations in failure rates. We first applied a chi-square test, which yields a p-value of 1.4879e-18 (< 0.05). This result indicates that LLMs' compliance performance depends on instruction intents rather than occurring independently, as it reflects systematic influences from their inherent characteristics.

We further examined the distribution of failure rates across instruction intent types, as shown in Table ~\ref{rq_2_intent}. Intent types such as questioning LLM responses, refactoring code, fixing bugs, and requesting code implementation exhibit high failure rates, exceeding 70\%, 60\%, 60\%, and 50\%, respectively. 
These intents typically involve code modifications, error corrections, or new functionality implementations, which require LLMs to handle complex constraints and contexts. In contrast, intents such as querying information and clarifying requirements show lower failure rates, approaching 45\% and 50\%. This pattern highlights the relative simplicity and determinism of query-based tasks, which facilitate accurate responses from LLMs. Intents such as adding new features and reviewing code fall in between, indicating moderate levels of challenge.

\input{tab/rq_2_intent}

Standardized residuals reveal distinct patterns in intent and compliance associations: Fix bugs intent shows significant bias toward non-compliance (residuals: -2.08 for compliance, 2.00 for non-compliance). Question LLM response intent approaches the significance threshold, trending toward non-compliance (residuals: -1.75, 1.68). Query information intent leans positively toward compliance (residuals: 1.65, -1.59), while refactor code and request code implementation intents exhibit notable negative deviations (absolute residuals >1.0).

In addition, we also observed instances of instruction forgetting in LLMs, where the forgetting step length is defined as the interval between the initial appearance of a specific instruction intent type and its first failure in a conversation. Intent types without any failures were excluded from this analysis. 
Through initial analysis and quantification, we identified 46 instances of instruction non-compliance attributable to the model's failure to retain prior-turn instructions in the current context.
Step lengths of 1 and 2 dominate the distribution, accounting for approximately 26\% and 33\%, respectively, for a combined share exceeding 58\%. 
Proportions decrease progressively from lengths 3 to 8, with those beyond 5 comprising just 4\%. 
Cumulatively, about 59\% of forgetting instances occur within 2 steps, rising to 87\% within 4 steps. 
The mean step length is 2.65 turns, suggesting that forgetting typically emerges within roughly 3 turns.

\begin{tcolorbox}[colback=lightgray, colframe=lightgray, boxrule=0.5pt, arc=5pt,left=5pt,right=5pt,top=5pt,bottom=5pt]
\textbf{Finding 6:}
Intents such as fixing bugs, refactoring code, and questioning LLM responses exhibit high rates of non-compliance and are prone to errors. In contrast, querying information demonstrates significantly better following of instructions.

\end{tcolorbox}

%% file: tab/rq_2_std.tex
\begin{table}[t!]
\setlength{\abovecaptionskip}{0.1cm}
\caption{Statistics of Non-Compliance by Interaction Patterns}
\renewcommand{\arraystretch}{1.1}
\setlength{\tabcolsep}{1.5pt}

\begin{tabular}{ 
    >{\centering\arraybackslash}m{1.6cm} 
    >{\centering\arraybackslash}m{1.6cm} 
    >{\centering\arraybackslash}m{1.6cm} 
    >{\centering\arraybackslash}m{1.6cm} 
    >{\centering\arraybackslash}m{1.6cm} }
\toprule
\textbf{Patterns} & \textbf{Count} & \textbf{Mean} & \textbf{Std} & \textbf{Non-compliance rate (\%)} \\ 
\midrule
Linear & 249 & 0.7349 & 0.4422 & 73.49 \\ 
Star & 59 & 0.6440 & 0.4829 & 64.40 \\ 
Tree & 70 & 0.9428 & 0.2337 & 94.28 \\ 
\bottomrule
\end{tabular}
\label{rq_2_std}
\vspace{-0.4cm}
\end{table}

%% file: tab/rq_2_turn.tex
\begin{table}[t!]
\setlength{\abovecaptionskip}{0.1cm}
\caption{Position Distribution and KS Test Results}
\footnotesize 
\renewcommand{\arraystretch}{1.1}
\setlength{\tabcolsep}{1.5pt}

\begin{tabular}{ 
    >{\centering\arraybackslash}m{2.5cm} 
    >{\centering\arraybackslash}m{2.5cm} 
    >{\centering\arraybackslash}m{1.5cm} 
    >{\centering\arraybackslash}m{1.5cm} }
\toprule
\textbf{Turn} & \textbf{Position} & \textbf{Position Distribution (\%)} & \textbf{KS Test} \\ 
\midrule
\multirow{4}{*}{Short Dialogues (turn $\leq$4)} & 1 & 38.68 & \multirow{4}{*}{1.2005e-31} \\ 
 & 2 & 41.22 &  \\ 
 & 3 & 14.25 &  \\ 
 & 4 & 5.85 &  \\ 
\midrule
\multirow{5}{*}{Long Dialogues (turn $>$4)} & Early (1-20\%) & 16.48 & \multirow{5}{*}{5.7985e-06} \\ 
 & Early-Mid (21-40\%) & 18.35 &  \\ 
 & Mid (41-60\%) & 17.98 &  \\ 
 & Mid-Late (61-80\%) & 19.85 &  \\ 
 & Late (81-100\%) & 27.34 &  \\ 
\bottomrule
\end{tabular}
\label{rq_2_turn}
\vspace{-0.4cm}
\end{table}

%% file: tab/rq_2_intent.tex


\begin{table}[t!]
\setlength{\abovecaptionskip}{0.1cm}
\caption{Instruction Following and Non-Compliance Rates by Type}
\small
\renewcommand{\arraystretch}{1.1}
\setlength{\tabcolsep}{1.5pt}

\begin{tabular}{ 
    >{\raggedright\arraybackslash}m{2.3cm} 
    >{\centering\arraybackslash}m{1.2cm} 
    >{\centering\arraybackslash}m{1.6cm} 
    >{\centering\arraybackslash}m{1.3cm} 
    >{\centering\arraybackslash}m{1.6cm} }
\toprule
\multicolumn{1}{>{\centering\arraybackslash}m{2.3cm}}{\textbf{Instruction Intent}} & \multicolumn{1}{>{\centering\arraybackslash}m{1.2cm}}{\textbf{Following}} & \multicolumn{1}{>{\centering\arraybackslash}m{1.6cm}}{\textbf{Non-Compliance}} & \multicolumn{1}{>{\centering\arraybackslash}m{1.3cm}}{\textbf{Following rate}} & \multicolumn{1}{>{\centering\arraybackslash}m{1.6cm}}{\textbf{Non-Compliance Rate(\%)}} \\ 
\midrule
Add New Feature & 23 & 35 & 39.66 & 60.34 \\ 
Clarify Requirement & 64 & 65 & 49.61 & 50.39 \\ 
Fix Bugs & 79 & 128 & 38.16 & 61.84 \\ 
Polite Greeting & 21 & 0 & 100.00 & 0.00 \\ 
Query Information & 186 & 156 & 54.39 & 45.61 \\ 
Question LLM Response & 12 & 29 & 29.27 & 70.73 \\ 
Refactor Code & 38 & 62 & 38.00 & 62.00 \\ 
Request Code Implementation & 137 & 179 & 43.35 & 56.65 \\ 
Review Code & 2 & 3 & 40.00 & 60.00 \\ 
\bottomrule
\end{tabular}
\label{rq_2_intent}
\vspace{-0.4cm}
\end{table}

%% file: sec/6_human.tex
\section{RQ3: Human Satisfaction }

\subsection{Methodology}
This section employs a dual-scoring mechanism to assess human satisfaction with the interaction process.
First, we used the deepseek-reasoner to assign a five-point satisfaction score to LLM responses within conversations. 
The scoring criteria draw from the WB-Score evaluation standard in WildBench, with scores ranging from 1 (highly unsatisfactory, with severe issues) to 5 (highly satisfactory, delivering an excellent experience that exceeds expectations). 
The evaluation process adopted default parameter settings. Second, to ensure the reliability of LLM-based scoring, we conducted manual reviews following the procedure outlined in Section 5.1. 
During these reviews, we observe that users incorporate positive or negative vocabulary in their instructions to express satisfaction or dissatisfaction with LLM responses. For instance, the presence of negation words such as ``don’t'', ``no'', or ``not'' in the current instruction often indicates dissatisfaction with the previous interaction. If users highlight errors or irrelevant information in the prior LLM response during the next turn, it suggests that the LLM fails to fully implement the user's instruction correctly. Conversely, the inclusion of expressions of gratitude in the subsequent instruction signals satisfaction with the preceding LLM response. Accordingly, we integrated assessments of LLM response quality with analyses of common emotional vocabulary to evaluate human satisfaction comprehensively. The manual review results align with LLM evaluations in over 87\% of cases, establishing a solid data foundation for subsequent quantitative analysis.

\subsection{Result and analysis }




\subsubsection{Task and Human Satisfaction}

\begin{figure}
    \centering
    \setlength{\abovecaptionskip}{0.1cm}
    \includegraphics[width=0.45\textwidth]{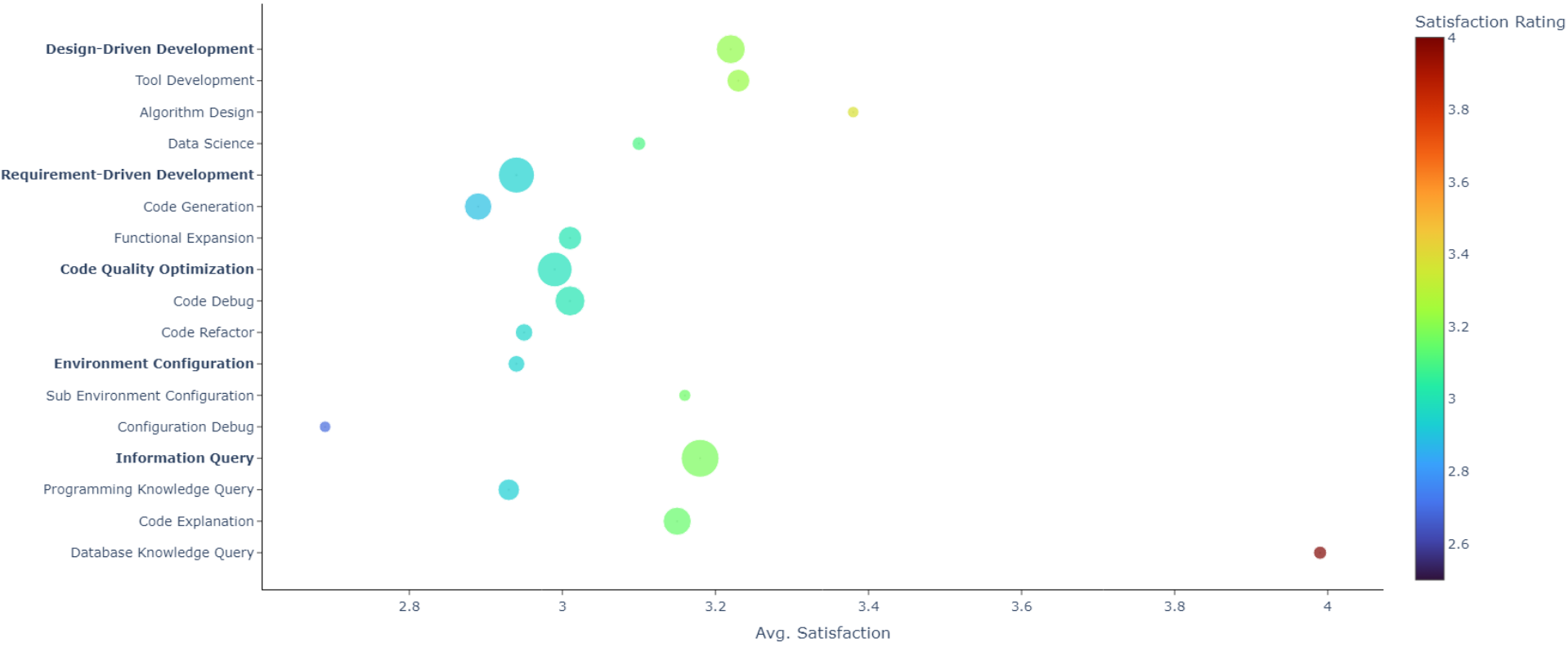}
    \caption{Average User Satisfaction Ratings Across Tasks and Subtasks}
    \label{rq_3_pao}
    \vspace{-0.7cm}
\end{figure}

This section analyzes user satisfaction with different task types to reveal user performance and its underlying patterns.
Figure ~\ref{rq_3_pao} illustrates the average satisfaction scores for five primary task categories and their subtasks. 
Analyses reveal substantial capability variations and uneven user experiences in how models address diverse development tasks.

First, from a macro-level view of task categories, user satisfaction correlates strongly with the software development phase and the abstraction level involved. 
Tasks emphasizing conceptual design and information planning in early development stages generally elicit higher user satisfaction: design-driven development (average score: 3.22) and information query (average score: 3.18) rank highest. 
Users in these contexts often treat the LLM as a knowledgeable advisor for framework design, direction exploration, or knowledge acquisition. 
By contrast, satisfaction diminishes as tasks advance to concrete coding implementation and environment setup. Categories such as requirement-driven development (average score: 2.94), environment configuration (average score: 2.94), and code quality optimization (average score: 2.99) all register lower scores. 
Here, users expect the LLM to produce directly runnable code or configurations.
Consequently, errors, incompatibilities, or inefficiencies in outputs obstruct progress and provoke negative feedback.

Second, in-depth subtask examination provides more concrete explanations and corroboration.
Database knowledge query emerges as the top-performing subtask, with an average score of 3.99. This pattern indicates that LLMs deliver precise, satisfying responses when users seek knowledge with well-defined structures and rules, such as database schemas or SQL syntax. In sharp contrast, configuration debug yields the lowest average score of 2.69 across all subtasks. Such outcomes expose a fundamental LLM limitation: performance falls short in diagnostic and repair scenarios that lack standardized patterns and demand context-specific, iterative trial-and-error. Within individual major categories, subtasks further display pronounced satisfaction imbalances. 
For information query, database knowledge query (score: 3.99) substantially outperforms programming knowledge query (score: 2.93). Likewise, under environment configuration, generative subtasks like environment configuration and deployment (score: 3.16) garner significantly higher ratings than reparative ones like configuration debug (score: 2.69). These findings underscore pervasive capability asymmetries in LLMs during problem resolution.

\begin{figure}
    \centering
    \setlength{\abovecaptionskip}{0.1cm}
    \includegraphics[width=0.45\textwidth]{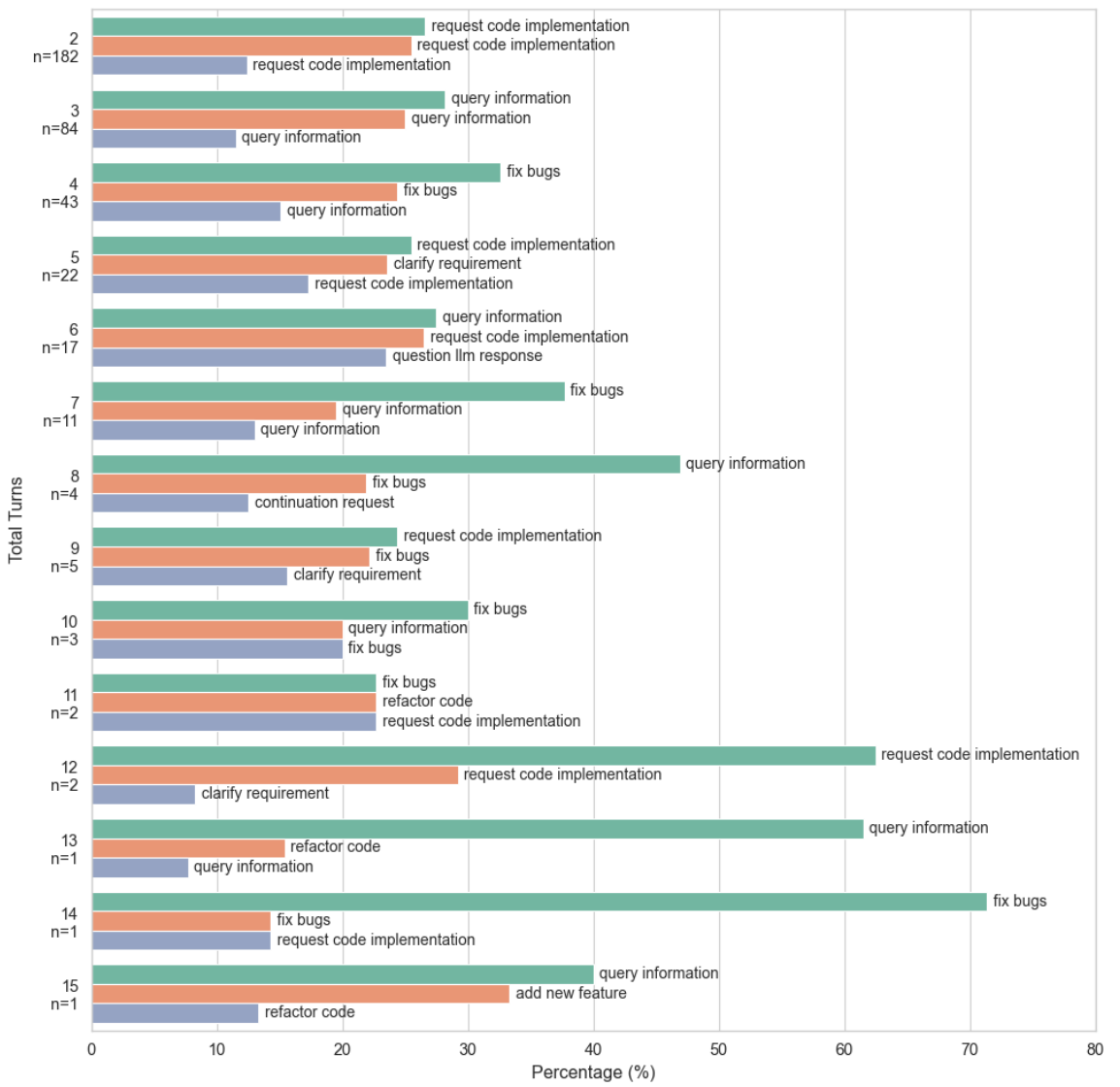}
    \caption{Distribution of Core Tasks Across Varying Conversation Lengths} 
    \label{rq_3_tiao}
    \vspace{-0.6cm}
\end{figure}

Building on these task-specific insights, we further examine how interaction dynamics influence satisfaction by analyzing core task distributions across conversations of varying lengths. 
As shown in Figure ~\ref{rq_3_tiao}, user satisfaction declines as dialogues extend, largely due to interactions devolving into inefficient "generate-revise" cycles that elevate the proportion of error-correction tasks.
In shorter conversations (2-4 turns), core tasks primarily involve constructive activities like code implementation and information querying, with bug-fixing comprising only 11.5\% to 15.1\% of tasks.
However, a shift occurs from the 5th turn onward, where bug-fixing surges to become a dominant activity—reaching 25.5\% in the 5th turn and 27.5\% in the 6th, often claiming the top spot.
In longer dialogues, bug-fixing maintains prominence, fluctuating in rank but peaking at 71.4\% in 14-turn conversations. 
Occasional resurgences of information querying (e.g., leading in the 7th, 12th, and 13th turns) typically reflect users rephrasing queries or seeking alternatives after suboptimal LLM outputs, rather than true progress.

In summary, conversation length extends in tandem with falling satisfaction, as interaction efficiency diminishes. Extended dialogues typically reflect initial LLM outputs that miss user needs. This forces additional turns for clarification, revision, or retries. The resulting move from one-shot resolutions to iterative debugging drives the satisfaction drop with increasing conversation length.

\begin{tcolorbox}[colback=lightgray, colframe=lightgray, boxrule=0.5pt, arc=5pt,left=5pt,right=5pt,top=5pt,bottom=5pt]
\textbf{Finding 7:}
Code quality optimization and requirements-driven development tasks show lower user satisfaction, whereas structured knowledge queries and algorithm designs yield higher levels.
Moreover, satisfaction declines as conversations lengthen, with task focus increasingly shifting toward error correction.


\end{tcolorbox}

\subsubsection{Interaction Pattern and Human Satisfaction}

This section examines the relationships between interaction patterns and user satisfaction. 
To capture dynamic shifts in user sentiment, analyses introduce the concept of satisfaction trajectories. 
The satisfaction progression of each conversation falls into distinct categories. Our analysis of user challenges mainly focuses on three particularly revealing trajectories: downward trends, which denote interactions with steadily declining satisfaction; trend reversals, which represent successful shifts from low to high satisfaction; and dual-turn lows, which specifically indicate failures concluding with low satisfaction across two turns.

Our analysis reveals that the tree pattern emerges as a primary source of poor user experiences. This pattern highlights limitations in current LLMs when handling complex, multi-path exploratory tasks. Interactions in the tree pattern extend the longest, averaging 5.7 turns per conversation. Yet this extended engagement yields diminishing returns, with final-turn average satisfaction reaching only 2.80—the lowest across all patterns. 
The underlying cause is further elucidated by the satisfaction trajectories: downward trends account for 17.1\% of cases in this pattern, the highest proportion observed.
Such prevalence suggests that LLM performance deteriorates progressively in intricate dialogues, as models struggle to manage multiple logical threads effectively. This dynamic often culminates in user disengagement.
The star pattern, despite shorter average turns, exposes deficiencies in LLMs' ability to deliver diverse and effective alternatives. A hallmark of this pattern lies in the elevated proportion of dual-turn lows, which reaches 30.5\%. These scenarios typically arise when users express dissatisfaction with initial LLM responses and pursue varied reformulations from the same starting point to elicit better options. Repeated failures, however, prompt abrupt conversation termination.

Analyses reveal that negative experiences pervade all patterns, with LLMs proving challenging to recover from adverse cycles once initiated.
For instance, dual-turn lows dominate as the chief negative form in brief two-turn dialogues. In contrast, downward trends prevail in extended three-to-five-turn exchanges. Notably, trend reversals remain rare across patterns, underscoring the difficulty LLMs face in rebounding from early setbacks to regain user satisfaction.


\begin{tcolorbox}[colback=lightgray, colframe=lightgray, boxrule=0.5pt, arc=5pt,left=5pt,right=5pt,top=5pt,bottom=5pt]
\textbf{Finding 8:}
The tree pattern emerges as a primary locus of poor user experiences, which exposes model limitations in managing complex tasks. 
Negative experience trajectories prevail across patterns, as models struggle to recover from early failures.

\end{tcolorbox}







%% file: sec/7_discussion.tex
\section{Discussion}
This section explores the implications of our findings for
key stakeholders—developers and researchers —and identifies the threats to the validity of our research.
\subsection{Implications}

\textbf{Individual Developers.}
\textbf{Leverage multi-turn interactions for iterative task refinement while prioritizing linear patterns for efficiency (Finding 1, Finding 2).} For developers using LLMs such as ChatGPT or Copilot in coding tasks, multi-turn conversations serve as a valuable mechanism to progressively refine outputs, such as debugging code or optimizing algorithms, as substantiated by our identification of linear, star, and tree patterns. However, our analysis reveals that linear patterns dominate low-turn scenarios (Finding 2), enabling rapid resolutions with minimal complexity. 
Developers should initiate interactions with clear, sequential prompts to promote linear flows, particularly for bug fixing or refactoring, where instruction non-compliance rates exceed 60\%. This strategy minimizes the risk of becoming disoriented in the conversation, a prevalent issue in multi-turn LLMs, as recent studies have highlighted, often stemming from compounding errors that degrade performance ~\cite{laban2025llms}. 
If a task escalates to higher turns (> 4), developers should monitor for shifts to tree patterns, which our findings identify as high-risk (94.28\% non-compliance rate), and intervene early by rephrasing prompts to preserve context.

\textbf{Researchers.}
\textbf{Developing automated methods for human satisfaction evaluation in multi-turn conversations warrants attention (Finding 7, Finding 8).}
 In this study, we employed a satisfaction trajectory approach to capture dynamic changes across multi-turn interactions. 
 Although we incorporated considerations of recency effects ~\cite{siro2022understanding,glanzer1966two} during manual scoring to the extent possible, the single-turn response scoring mechanism introduces limitations, potentially overlooking the non-linear accumulation of emotions.
This limitation prompted us to survey existing human satisfaction evaluation methods, revealing that most approaches are grounded in single-turn assessments.
For embedded methods ~\cite{deng2022user,ye2023modeling}, common techniques include leveraging BERT or similar embedding models to compute semantic similarity scores between responses and user intents, thereby quantifying consistency and relevance.
In contrast, LLM-based methods ~\cite{kim2025llm,lin2024interpretable} typically rely on zero-shot or few-shot learning to guide an LLM as an evaluator.
While these methods effectively capture single-turn quality, they fall short in reflecting cumulative emotional evolution in multi-turn conversations.

To further validate this hypothesis, we randomly selected 112 multi-turn conversations. 
Due to the absence of overall conversation satisfaction information, we made every effort to assign an overall score to each conversation by applying the recency effect from psychology. 
The scoring process was primarily conducted by a team of student annotators, followed by a consistency evaluation of the annotation results. 
The evaluation findings revealed an inter-annotator agreement exceeding 70\%
We then applied zero-shot and few-shot learning (with three examples) to prompt GPT-3.5-turbo and GPT-4o for scoring. Additionally, we utilized the SPUR framework, which operates in few-sample scenarios by employing iterative prompting to generate evaluation rubrics from conversations and subsequently guide LLM-based satisfaction assessments.

\begin{figure}
    \centering
    \setlength{\abovecaptionskip}{0.1cm}
    \includegraphics[width=0.45\textwidth]{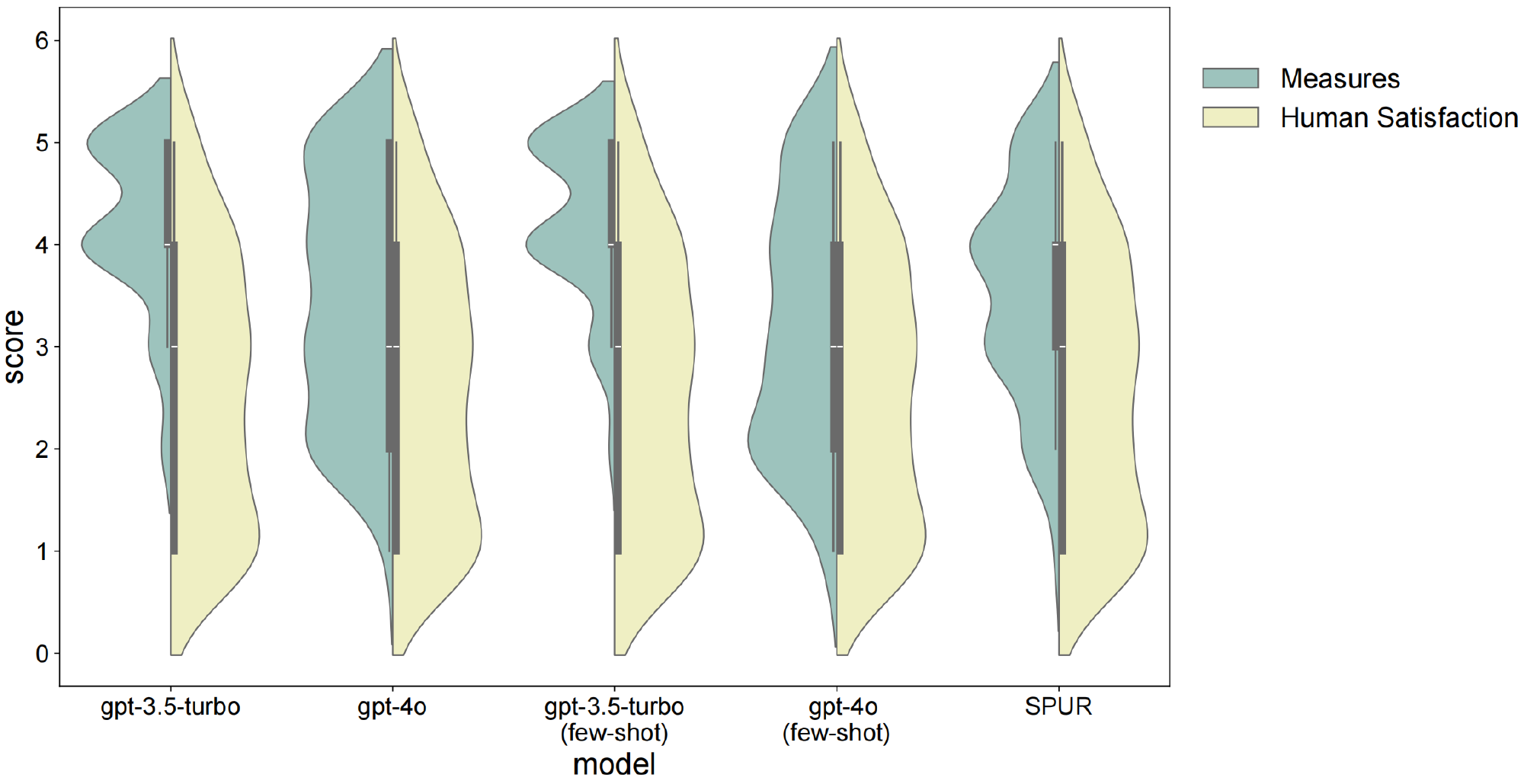}
    \caption{Value distribution of the four different measures and human satisfactions} 
    \label{manyidu_fig}
    \vspace{-0.6cm}
\end{figure}

Figure ~\ref{manyidu_fig} illustrates the distributional differences between these methods and human satisfaction scores. Preliminary results show that LLM-based evaluations tend to overestimate human satisfaction optimistically, while the SPUR method, reliant on extracting rubrics from conversations, exhibits biases on our dataset, potentially due to data-specific characteristics. These findings underscore that current evaluation methods inadequately capture human satisfaction in multi-turn contexts.
Future researchers could integrate psychological factors—such as recency effects ~\cite{siro2022understanding,glanzer1966two} and the non-linear amplification ~\cite{feng2024infusing,song2022emotionflow} of negative emotions—to assess human satisfaction in multi-turn interactions. This is essential for unlocking the true potential of human-LLM collaboration and fostering efficient, harmonious, and sustainable human-LLM partnerships.

\subsection{Threats to Validity}
\textbf{External Validity.}
While our analysis draws from coding-related dialogues in the LMSYS-Chat-1M and WildChat datasets, which largely reflect interactions with widely used LLMs such as ChatGPT and encompass a diverse array of user populations across various contexts, the applicability of these insights to other models remains an area for further exploration. Although these datasets span a broad spectrum of subjects, their focus on specific platforms may influence the observed patterns. Moreover, the selection of 378 multi-turn exchanges provides a targeted perspective on coding activities, which might not fully capture the evolving capabilities of LLMs that have advanced since the data collection period.

\textbf{Internal Validity.}
Employing DeepSeek to assess adherence to instructions and user contentment could introduce subtle discrepancies, given that its analytical strengths, while robust, may not always mirror human evaluations precisely—even with targeted manual checks on 20\% of the samples. Annotations conducted by domain experts and graduate students demonstrated strong consistency (Cohen's Kappa exceeding 0.78), though individual perspectives on task categories, dialogue flows, and underlying motivations could subtly color the interpretations. We utilized appropriate statistical methods, including Kruskal-Wallis and Fisher's exact tests, for handling non-normal distributions; however, elements like differing conversation durations, if not fully addressed, could influence the strength of our conclusions regarding causality.

\textbf{Construct Validity.}
Our reliance on LLM-based evaluations for metrics such as non-compliance rates and satisfaction scores may not fully capture the subtle dynamics in human-LLM interactions or implicit user intents. For instance, satisfaction is inferred from response patterns and explicit feedback, which could overlook finer indicators like conversation abandonment. 
Moreover, the satisfaction assessment might not entirely account for psychological factors, such as the recency effect. 




%% file: sec/8_conclusion.tex
\section{Conclusion}
\label{sec:conclusion}
In this study, we conduct an empirical analysis of human-LLM coding collaboration using LMSYS-Chat-1M and WildChat datasets, to explore the human-LLM collaboration mechanism, LLMs instruction-following ability, and human satisfaction. Our study has both theoretical and practical implications. 
By analyzing the empirical results, we enable a deeper understanding of the dynamics between users and LLMs in coding tasks, including interaction patterns, instruction following challenges, and satisfaction factors. 
We observe how task types influence interaction patterns—such as linear flows in code optimization, tree-like branches in design-driven tasks, and star configurations in queries—while interaction length and complexity drive shifts toward more intricate patterns, higher non-compliance in areas like bug fixing, and degraded experiences in prolonged exchanges marked by error correction and poor recovery in tree structures. 
Based on these findings, we provide recommendations for enhancing LLM interfaces to better support adaptive flows, improve instruction following, and boost user satisfaction in coding collaborations, as well as shed light on future research directions. 

